# Effects of Alloying Elements on Surface Oxides of Hot–Dip Galvanized Press Hardened Steel

Wolfgang Gaderbauer[1,2,*], Martin Arndt[3], Tia Truglas[1], Thomas Steck[3], Nico Klingner[4], David Stifter[5], Josef Faderl[3], Heiko Groiss[1]

*1 Christian Doppler Laboratory for Nanoscale Phase Transformations, Center for Surface and Nanoanalytics, Johannes Kepler University Linz, Altenberger Straße 69, 4040 Linz, Austria*

*2 K1–MET GmbH, Stahlstraße 14, 4020 Linz, Austria*

*3 voestalpine Stahl GmbH, Voestalpine Straße 3, 4031 Linz, Austria*

*4 Helmholtz–Zentrum Dresden–Rossendorf e.V., Institute of Ion Beam Physics and Materials Research, Bautzner Landstr. 400, Dresden 01328, Germany*

*5 Center for Surface and Nanoanalytics, Johannes Kepler University Linz, Altenberger Straße 69, 4040 Linz, Austria*

## Abstract

Effects of steel alloying elements on the formation of the surface oxide layer of hot–dip galvanized press hardened steel after austenitization annealing were examined with various advanced microscopy and spectroscopy techniques. The main oxides on top of the original thin $Al_2O_3$ layer, originating from the primary galvanizing process, are identified as ZnO and $(Mn,Zn)Mn_2O_4$ spinel. For some of the investigated steel alloys, a non–uniform, several nanometer thick Cr enriched, additional film was found at the $Al_2O_3$ layer. At a sufficiently high concentration, Cr can act as a substitute for Al during annealing, strengthening and regenerating the original $Al_2O_3$ layer with $Cr_2O_3$. Further analysis with secondary ion mass spectrometry allowed a reliable distinction between ZnO and $Zn(OH)_2$.

---

[*] Corresponding author: email: wolfgang.gaderbauer@k1-met.com

# 1. Introduction

In recent years, lightweight ultra–high strength steels (UHSS) have drawn increasing attention. With challenges such as emission reduction, stringent safety requirements and the new focus on e–mobility, ever-lighter construction of structural car body parts is essential for low-cost and high-value manufacturing. Hot–dip galvanized press hardened steel (PHS) provides excellent galvanic corrosion resistance, while maintaining the mechanical capabilities of hot–formed ultra–high strength steels [1,2].

Sagl et al. [3,4] investigated the role of oxidized alloying elements on the wetting behavior of Zn during galvanization and found them to be a major factor on the general galvanizability of the uncoated steel surface. In further investigations by Arndt et al. [5] the huge influence of pre-oxidized alloying elements (mainly Mn) on the wetting behavior could be confirmed and a detailed model for the wetting process was given.

In principle, Zn coatings for PHS are either continuously hot–dip galvanized (GI) zinc (Zn) coatings with low Al additions (< 0.5 wt.%) or galvannealed (GA) zinc–iron (ZnFe) coatings with even less Al (< 0.16 wt.%) added to the Zn bath. During hot–dipping in the liquid Zn bath, Al reacts with Fe from the steel strip surface and forms a $Fe_2Al_5$ inhibition layer. In Zn coatings, this layer prevents interdiffusion of Fe and alloying elements from the steel and the Zn coating. In ZnFe coatings however, the lower amount of Al in the Zn-bath results in no detectable inhibition layer during inductive heat treatment, where the steel strip with Zn on top is heated up to 550 °C [6]. During this galvannealing process, interdiffusion of Zn and Fe results in a coating with 8–12 wt.% Fe [7].

A detailed overview of the coating evolution during annealing is given by Kang et al. [8] and Järvinen et al. [9,10]. They performed microstructure analysis of Zn and ZnFe coatings on the steel–coating interface region for different manganese–boron steel grades. They found that higher C contents, as well as higher contents of the alloying elements Mn and Cr result in more stable γ–Fe(Zn) but the evolution of the surface oxide region was not investigated in their work. Chen et al. [11] focused on the effects of surface oxides on the steel strip before galvanization and compared standard Zn coatings on dual phase steels with ZnFe galvannealed coatings. Wang et al. [12] described the diffusion process for the ZnFe phase depending on the Zn concentration in GA coated 22MnB5.

Autengruber et al. [13] extended their previous work on Zn coated PHS and gave a brief overview of the post annealed coating surface. In their investigations, they found a mixture of mainly ZnO and $Mn_3O_4$ on top of a thin $Al_2O_3$ layer. The $Al_2O_3$ layer is a consequence of low Al additions in the Zn bath during galvanization and acts as a barrier, which prevents oxidation of the liquid Zn immediately after hot–dipping [6]. Based on these findings, Lee et al. [14] performed heating experiments and described the sequence of the oxide formation as a consequence of the temperature change during annealing. They showed that low heating rates correlate with a higher fracturing of the initial $Al_2O_3$ layer and thus with increased ZnO formation. Chang et al. [15,16] thoroughly described the microstructure evolution during austenitization, as well as the oxidation and corrosion behavior of 5 wt.% Al–Zn coated steel.

The results of these works indicate a strong influence of the alloying elements on the final coating structure and especially on the oxide formation. The quality of post–annealing processing techniques like spot welding, adhesive bonding, painting or application of further (organic) coatings are influenced by the oxides on the surface of the coated steel. Therefore, we investigate the uppermost oxide layer after press–hardening austenitization annealing of four industrial steel grades commonly used for PHS applications.

In order to get a complete picture of the surface oxide distribution a correlative characterization [17] was pursued. Beginning with optical light microscopy (OLM), first differences in the structural and optical appearance of the surfaces are investigated on mm to µm scale. A noticeable difference was the different visual appearance of the surface for GI and GA coated specimen. Due to the limited resolution of OLM further investigations with higher lateral resolution were necessary. By means of scanning electron microscopy (SEM) combined with additional detection techniques like energy dispersive X–ray spectroscopy (EDX) and Auger electron spectroscopy (AES) the prepared cross sections are investigated on a scale ranging from several µm down to sub µm. From these SEM results, a detailed overview of the phase mixture and microstructure of the coatings can be determined. Additional chemical investigations were made by time–of–flight secondary ion mass spectrometry (ToF–SIMS) powered by a helium ion microscope (HIM). Small features with only a few nanometers in diameter were found and their investigation required the sub-nanometer resolution of a transmission electron microscope (TEM). Using conventional TEM, it is possible to perform crystallographic analysis with selected area diffraction (SAD). EDX measurements in scanning mode (STEM) allowed for a reliable description of the different oxide phases in the coating.

## 2. Material and Methods

### 2.1. Sample material and preparation

The different elemental compositions of the investigated steel substrates are given in Table 1. With the exception of HX340LAD, all basic steel grades are manganese–boron steels with slight differences in alloy compositions. HX340LAD consists of less C, Mn and no B compared to the other specimens. 22MnB5 contains a high amount of Mn as well as significant amounts of Cr and Si. 20MnB8 is similar, with slightly higher Mn but without the additional Cr as part of the alloy composition. 22MnCrB8–2 shares the same high Mn content as 20MnB8 but has an additional low Cr content added, similar to the 22MnB5 specimen. In addition to the chemical differences of these steel grades, the Zn–coating types are varied. HX340LAD and 22MnB5 receive a standard hot–dip Zn coating (GI), while 20MnB8 and 22MnCrB8–2 are ZnFe coated (GA).

A continuously hot–dip galvanizing process includes a strip surface cleaning step and a recrystallization annealing step at temperatures from 700 to 850 °C in a radiant tube furnace under HNX atmosphere. During the heat treatment, selective oxidation within the steel matrix occurs, which defines the wettability of the steel strip surface. After hot–dipping in the liquid Zn bath at about 450 °C, excess Zn is removed by gas knives. If the coating received the additional galvannealing heat treatment, the coated steel strip enters the induction furnace right after hot–dipping and wiping. The desired temperature of about 500 °C is reached within seconds, which allows the transformation of a pure Zn coating to a ZnFe coating with 8 to 12 wt.% Fe. Subsequently, the galvanized steel strip is cooled for a complete solidification.

**Table 1 Alloy contents of the different steel grades in weight percent. All specimen have a sheet thickness of 1.5 mm.**

| Steel grade | Coating | C | Si max | Mn max | Al | Cr max | Ti+Nb max | B |
|---|---|---|---|---|---|---|---|---|
| HX340LAD | GI 70/70 | ≤ 0,11 | 0,5 | 1,4 | ≥ 0,015 | 0,05 | 0,10 | – |
| 22MnB5 | GI 70/70 | 0,20 – 0,25 | 0,5 | 2,0 | 0,02 – 0,10 | 0,50 | 0,05 | 0,002 – 0,005 |
| 20MnB8 | GA 90/90 | 0,17 – 0,23 | 0,5 | 2,5 | 0,02 – 0,30 | 0,05 | 0,05 | 0,002 – 0,005 |
| 22MnCrB8–2 | GA 90/90 | 0,20 – 0,25 | 0,5 | 2,5 | 0,02 – 0,30 | 0,50 | 0,05 | 0,005 |

For our investigation sheets of (297 × 210) mm² were cut from the coated steel strips and austenitization annealed in a Nabertherm N41/H lab oven with a final annealing temperature of 910 °C (see Fig. 1) in ambient atmosphere. Two annealing series have been produced, were the first has 45 s holding time and the second has 200 s holding time after reaching a temperature of 870 °C. The heating rates, which are determined via a heat couple attached to the surface of the specimen, for the GI and GA coated specimens are highly different due to their different compositions and heat absorption capabilities. The GI coated specimen (blue lines in Fig. 1a,b) had a rather constant heating rate of 5 °C s$^{-1}$ up to about 550 °C, with a quick rise to 670°C with 7 °C s$^{-1}$. The GA coated specimen (red lines in Fig. 1a,b) had a very high initial heating rate of up to 15 °C s$^{-1}$, which declined rapidly after reaching 670 °C and showed similar heating rates as the GI coated specimen above 670 °C.

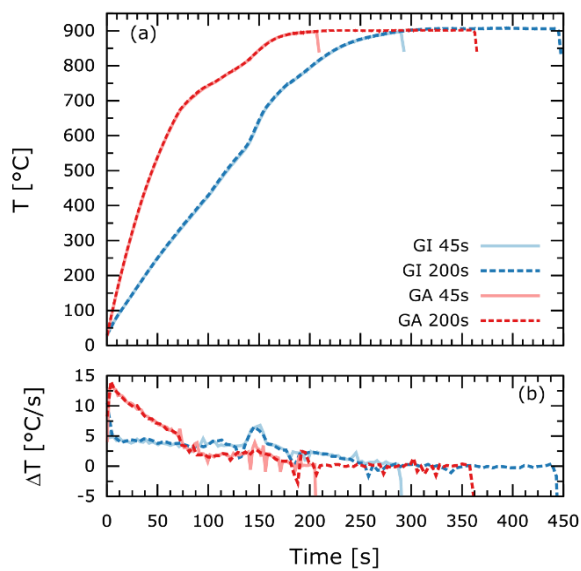

**Figure 1 Temperature curves measured with thermocouples during annealing in a lab oven for GI and GA coated specimens (a) and their corresponding heating rates (b).**

The heat treatment resulted in a phase–evolution of the Zn–Fe binary system as described by Marder [6,18] and Janik et al. [7]. Above 550 °C, the emissivity of the surface increases due to the replacement of liquid Zn with Zn–Fe crystals (δ–phase). At 670 °C the phase mixture is made of Γ–FeZn and liquid Zn, resulting in lower overall emissivity and thus lower heating rate. After the heat treatment, samples were quenched in ambient air, with cooling rates above 20 °C s$^{-1}$.

Small pieces with the dimension of (8 × 5) mm² were cut from the hardened steel plates. Before each measurement, each specimen was cleaned in an ultrasonic bath in various solvents (ethanol, 99.9%; acetone 99.9%; isopropanol 99.9%; tetrahydrofuran 99.9%) to remove surface contaminations resulting from the production process and sample handling. This is especially necessary for AES investigations due to the very high surface sensitivity, as similar investigations on ZnMgAl coated specimen have shown [19].

Sample preparation of focused ion beam (FIB)–thinned TEM lamellas is a challenging process due to the complex morphological structure and heterogeneous phase mixture of the coating. After austenitization heat treatment, large cavities underneath the porous and sometimes loose oxide layer had a negative influence on the stability of the lamellas during FIB thinning. Those structural factors resulted in milling artifacts, leading to non–uniformly thinned TEM–specimen [20]. Moreover, redeposition of the sputtered material onto the lamella was highly increased along these holes, creating artificial material mixtures, which may cover interesting features of the coating.

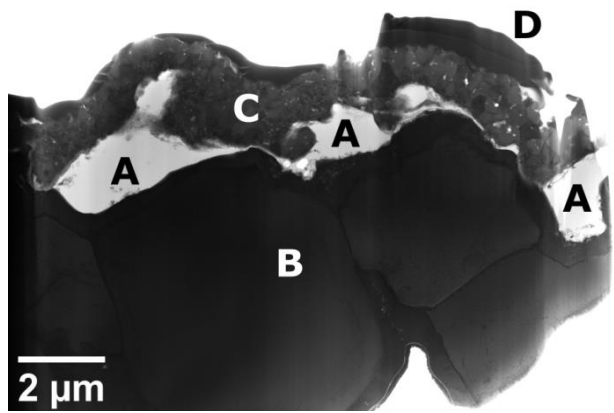

**Figure 2 STEM–BF image of a finished TEM–lamella where adhesive (A) was used to fill the pores, Zn–Fe intermetallic phases (B), oxide layer (C) and partially removed platinum cover (D).**

The best working solution found, was an application of the two–component, solvent–thinned epoxy–phenolic adhesive M–Bond 610 from Vishay Micro–Measurements, which we typically use for conventional TEM–sample preparation. After the described cleaning process, a small droplet of the liquid adhesive with at least 2 mm diameter was applied on the surface. The droplet size determines the quality of the cavity filling, as it must provide enough volume to reliably reach deeper holes within the coating. Afterwards, the adhesive was cured at room temperature in a vacuum chamber for 48 hours. In vacuum and additionally by capillary effects, the adhesive was pulled into the holes through cracks and channels in the coating.

A sample on which this adhesive application was used is shown in Fig. 2. Here, the cavities filled with the adhesive (A) are located between the intermetallic Zn–Fe phases (B) and the oxide layer (C). On top of the oxide, one can see a partially remaining platinum layer (D) from the FIB—milling process. The now hardened epoxy adhesive mechanically supports the porous, fragile oxide layer. Moreover, redeposition at the edges of these cavities is prevented during FIB milling and no increased curtaining around the holes can be found.

Cross sections for SEM and helium ion microscope (HIM) investigations were prepared by manual mechanical grinding with SiC grinding paper and polishing with diamond paste with a grain size down to 1 µm. After cleaning, the final polishing was done by $Ar^+$ sputtering in a cross section polisher (CSP) Leica EM TIC 3X with 8 kV accelerating voltage at –100 °C. The sputtering process at cryogenic temperatures with three converging aligned ion guns reliably produced a smooth, hardly damaged surface for AES, EDX and SIMS measurements. In order to protect the coating from redeposition during $Ar^+$ sputtering a protective layer had to be applied. For this purpose, either the epoxy adhesive was sufficient or an additional ink layer from a black felt tip pen was applied. The prior application of the epoxy adhesive proved to be superior to the black pen, as it smoothened the cracks and cavities in the cross sections, resulting in almost no curtaining effects of the $Ar^+$ sputtering.

### 2.2. Analytical methods

In the scope of this work, several electron and field ion microscopes have been used. A Zeiss Supra 35 was used for EDX investigations with an X–Max$^N$ 80mm$^2$ detector from Oxford instruments on cross section polished specimen.

Auger electron measurements were performed in an ultra–high vacuum (UHV) Jeol JAMP 9500F field emission Auger electron spectroscope. The Auger microprobe uses a hemispheric analyzer, which provides a spectral range from 0 to 2500 eV. The instrument is operated with a primary electron energy of 30 keV and currents of 10 to 20 nA and supports recording of scanning Auger elemental mappings in constant analyzer energy mode. Additionally, an $Ar^+$ ion gun is available for sputtering of the specimen surface with ion energies of 0.2 to 3 keV. Elemental mappings are recorded first for carbon, followed by oxygen and afterwards the other elements according to their atomic number. This procedure ensures reliable C and O mappings, due to higher C contamination with increasing measurement duration.

A Zeiss 1540XB CrossBeam was used for imaging of the original sample surfaces. Additionally, the FIB column of the Zeiss 1540XB was used for TEM sample preparation, where an accelerating voltage of 30 kV and milling currents from 100 pA to 20 nA, were used for imaging and sample milling of the investigated specimen.

TEM measurements were performed on a Jeol JEM–2200FS, where a Schottky Field emission gun was operated at 200 kV acceleration voltage. The microscope can be used in standard TEM or in scanning TEM mode (STEM). In STEM mode, a bright field (BF) detector as well as a high angle annular dark field (HAADF) detector were used. Attached to the microscope column is a silicon drift detector X–Max$^N$ 80T from Oxford Instruments for nanoscale EDX investigations.

The helium ion microscope (HIM) Orion NanoFab by Zeiss, was operated at 30 keV with neon as primary ion source and attached with a custom designed ToF–SIMS setup [21]. The high brightness of the ion source can reach up to $10^9$ A cm$^{-2}$ sr$^{-1}$ and allows for a lateral resolution for sputtering of 1.8 nm. The sample is biased at ±500 V in order to select either positive or negative ions for the ion spectrometer [21,22].

## 3. Results

### 3.1. Cross section analysis with SEM–EDX

Detailed investigations with SEM–EDX were made on CSP samples with the pre–applied adhesive stabilization (see Figure 3). The secondary electron (SE) image depicted in Figure 3a already shows a complex multi–phase structure according to the different gray values arising from Z–contrast. The mapping shown in Fig. 3b contains Fe, Zn and Mn and reveals a clear separation of at least two Zn–Fe intermetallic phases (1) and Zn– and Mn–rich oxides on top. In the mapping showing the remaining alloying elements (see Figure 3c), in between the oxide and Zn–Fe phases, a thin and slightly fractured $Al_2O_3$ layer can be identified (2). Alongside the $Al_2O_3$ layer, small particles of $SiO_2$ can be found (3). In the top right corner, above the porous oxide layer, the applied epoxy adhesive is clearly visible as it consists mainly of C.

In order to achieve a sufficiently high lateral resolution to reliably locate the small $Al_2O_3$ particles of the layer, a primary electron energy of 5 keV was used. A downside of the low excitation energy is the restricted spectral response, limiting the detectable spectral lines to low energy X–rays. This constraint affected the measurement adversely while monitoring Cr combined with O, as the Cr Lα line is overshadowed by the strong O Kα line. Therefore, the Cr signal cannot be separated from the O signal and the measured Cr distribution is highly related to O.

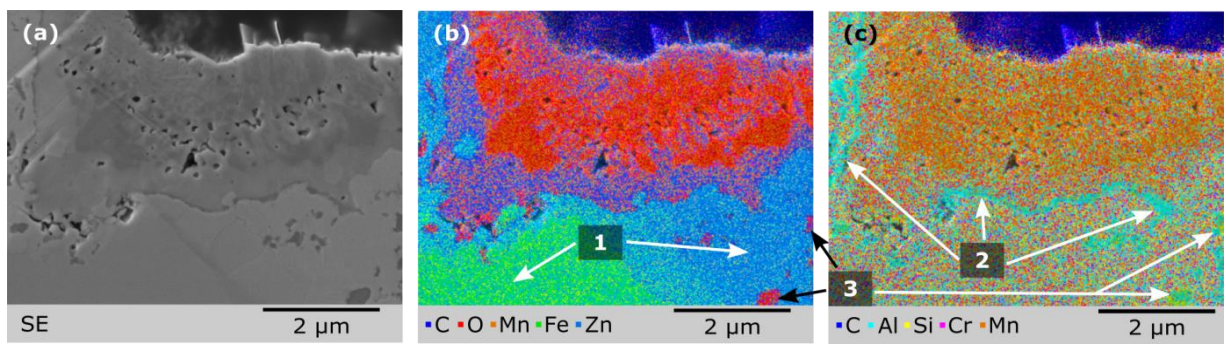

**Figure 3 SEM image and EDX mappings of a CSP 22MnCrB8–2 + GA (45 s > 870 °C) sample; (a) SE image; (b,c) EDX mappings with 5 keV primary electron energy. The layered structure shows two distinctive Zn–Fe intermetallic phases (1), which are separated from the Mn and Zn–rich oxide by a thin $Al_2O_3$ layer (2). Within the Γ–phase, small particles of Al and Si oxide can be found (3).**

## 3.2. Cross section analysis with AES

Complementary to the SEM–EDX measurements, scanning AES is also used for elemental mappings of cross sections. A major advantage of AES over EDX is the low information depth and interaction volume as well as a high lateral resolution. However, the high surface sensitivity requires special care while cleaning the samples. The prior application of the epoxy adhesive results in an ever increasing surface carbon layer during measurements due to electron beam induced deposition and surface diffusion processes.

Figure 4 shows mappings on the interface between oxide and Zn–Fe coating. Similar to the SEM–EDX measurement, the SE image shows various phases, easily distinguishable by their respective gray values in the image. In the center of the images, higher concentrations of C, O and Al are present in the Zn–Fe phases. These indicate residuals from the partially fractured, primary $Al_2O_3$ layer. This layer is the separator between oxide and intermetallic Zn–Fe phases and on a close look it is also visible as missing intensity in the Zn mapping. Similar to the previously shown SEM–EDX mappings (see Fig. 3), the top oxide layer consists of a Mn–rich oxide phase, which is enclosed in a Zn–rich oxide. The Fe and Zn mappings allow a clear distinction between the Fe–rich $\alpha$–Fe(Zn) and the Zn–rich $\Gamma$–$FeZn_7$. The effect of carbon contamination can be seen in the C–mapping. The top, oxide area has a strongly growing C layer, originating from the epoxy (out of view). Notable observations are the higher concentration of C on the $\alpha$–Fe phase at the bottom and on the small $Al_2O_3$ particles embedded in $\Gamma$–phase. We assume, this behavior originates from a higher contamination of these areas with C.

The same cross section polished specimen was further investigated with the Ne powered HIM. The results of a ToF–SIMS measurement depicted in Fig. 5 were recorded using a positive sample bias while the results shown in Fig. 6 were recorded with a negative bias. The mass to charge ratio is calibrated and calculated using known element peaks within the spectrum and assigning them to their respective elements like H, F and Al. This mapping allows a reasonable allocation of all spectral peaks to most elements, isotopes and small molecules.

The SE image in Fig. 5a shows the general structure of the coating layer on a large scale. One can see the epoxy adhesive as dark regions on top of the specimen and as a filling in the cavity between the oxide layer and intermetallic Zn–phases. The oxide layer is visible as a distinct layer running from top left to middle right. In the bottom area, a single large $\alpha$–Fe(Zn) grain is covered by $\Gamma$–$ZnFe_7$ phase. On the right–hand side, the oxide layer is attached to the Zn–Fe phases. The strongly varying gray values within the different structures are attributed to the different crystal orientations, leading to channeling and different SE yields in the HIM [23]. Diagonally running scratches can be seen in the center of the image. These are artifacts from the cross section polishing process, but due to the prior applied epoxy those scratches are not deep and thus negligible for the ToF–SIMS measurements.

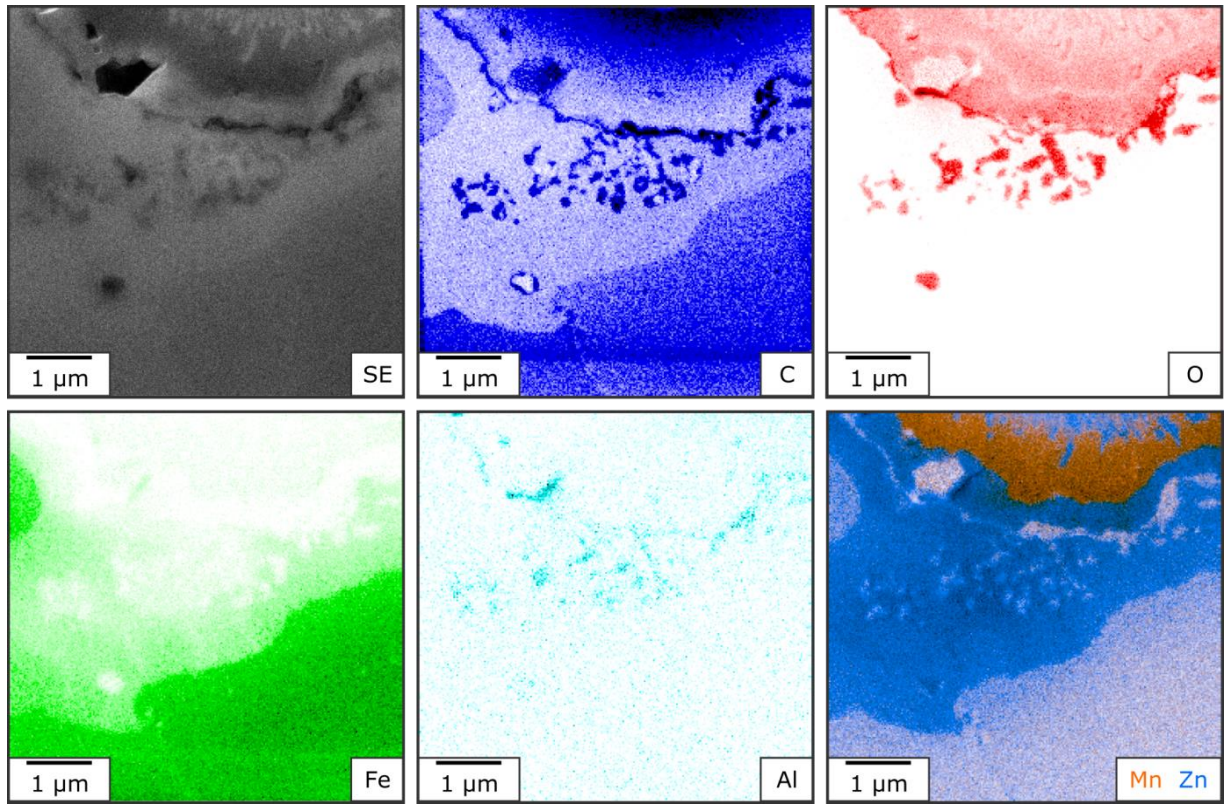

**Figure 4** AES mapping on a cross section of a 22MnCrB8–2 + GA (200 s > 870 °C) sample. The elemental distribution shows the structure of the finalized coating. Mn oxide is embedded in Zn oxide. The $Al_2O_3$ layer is highly fractured but lies also in between oxide and Zn–Fe intermetallic phases. The C signal is more intense in the areas of the $Al_2O_3$ particles, caused by surface diffusion from the epoxy adhesive. Darker colors correspond to higher intensity.

### 3.3. Cross section analysis with HIM–ToF–SIMS

In positive bias mode, metals have high relative sensitivity factors (RSF), as seen in the spectral response in Fig. 5d. Four peaks of the sum spectrum were selected, namely Al, Cr, Mn and Zn, and their respective elemental mappings are presented in Fig. 5b,c. As expected from the EDX and AES measurements, the uppermost oxide layer consists mainly of Zn with a distinct Mn enriched area embedded into Zn on the right. The $Al_2O_3$ film from the galvanization step is located on the bottom side of the ZnO layer. Additionally, Cr seems to accompany Al and even expanding the thin oxide film, where the Al response declines. Both elements together form a continuous layer on the bottom side of the top Mn and Zn layer.

A closer look at the sum spectrum in Fig. 5d shows more prominent element and molecule peaks. At first, one can identify a strong H peak. Next, C can be seen with a much lower count rate. Similar to H, C is detected all over the investigated area and often accompanied by H to form various hydrocarbon molecules ($C_xH_y$). $^{19}F$ has a similar low response as carbon but can be identified clearly, as there are no other isotopes with a similar mass–to–charge ratio. Beside the distinct peak of $^{27}Al$, $^{55}Mn$ has the highest yield of the detected metals. Right before $^{55}Mn$, one can see a small peak of $^{52}Cr$, which is almost overshadowed by the strong Mn peak. The next prominent peak is $^{64}Zn$ before one can see another peak at 71 u, which is assumed to originate from MnO.

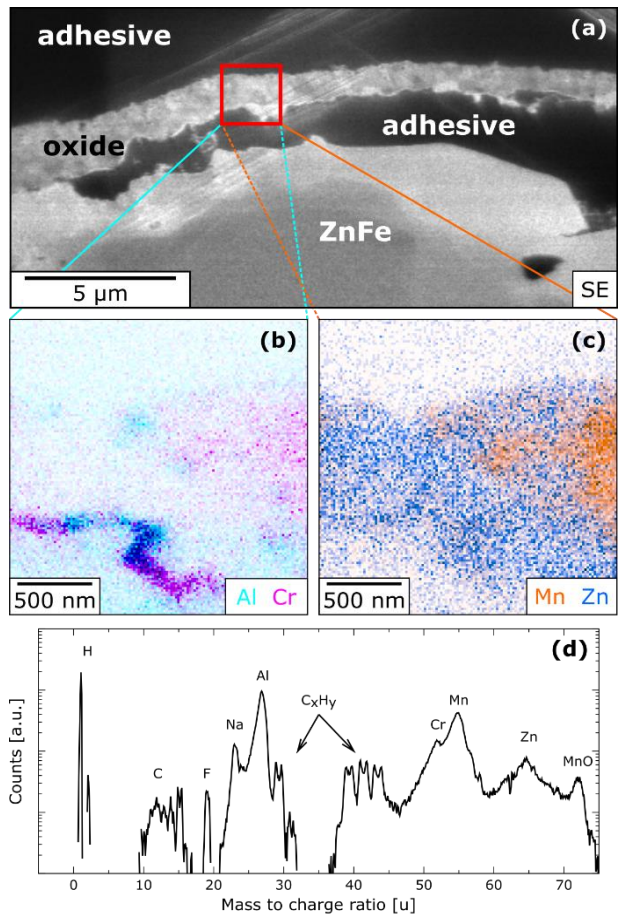

**Figure 5** Positively biased ToF–SIMS measurements on a cross section polished and epoxy stabilized 22MnCrB8–2 + GA (200 s > 870 °C) sample with SE image (a), elemental mappings of Al, Cr, Mn and Zn (b,c) and the sum spectrum of the measurement (d).

One of the measurements on a negatively biased specimen is shown in Fig. 6. The layered structure is once more visible in the SE image in Fig. 6a, but the overall image quality is comparatively low, due to the used fast image acquisition procedure to avoid sample damage during imaging. For the highlighted area, the mappings of elemental O and of OH are depicted in Fig. 6b,c. At a first glance, O and OH seem similarly distributed and can be found in the topmost layer of the coating. On a closer look however, the mappings differ in the bottom region, where O shows a loose cluster of smaller particles in the left half, while OH is highly localized at the bottom right quarter.

The sum spectrum for the negatively biased measurement is displayed in Fig. 6d. One can identify mostly non–metallic elements beginning with a prominent H peak. The next notable spectral peaks are related to $^{12}$C and slightly stronger CH, followed by almost equal $^{16}$O and OH peaks. A rather strong $^{19}$F peak is similarly distinctive as in Fig. 5d. Some other peaks with higher mass–to–charge ratio are present, but most cannot be clearly assigned. The strongest peaks may be C$_2$H and $^{35}$Cl and the small peak at 32 u could be related to O$_2$.

The different yields of the detected elements are directly related to the RSF of said elements [24,25]. As RSF values are not yet available for He or Ne as primary ion sources, the measurements yield only qualitative information about the elemental composition of the investigated areas.

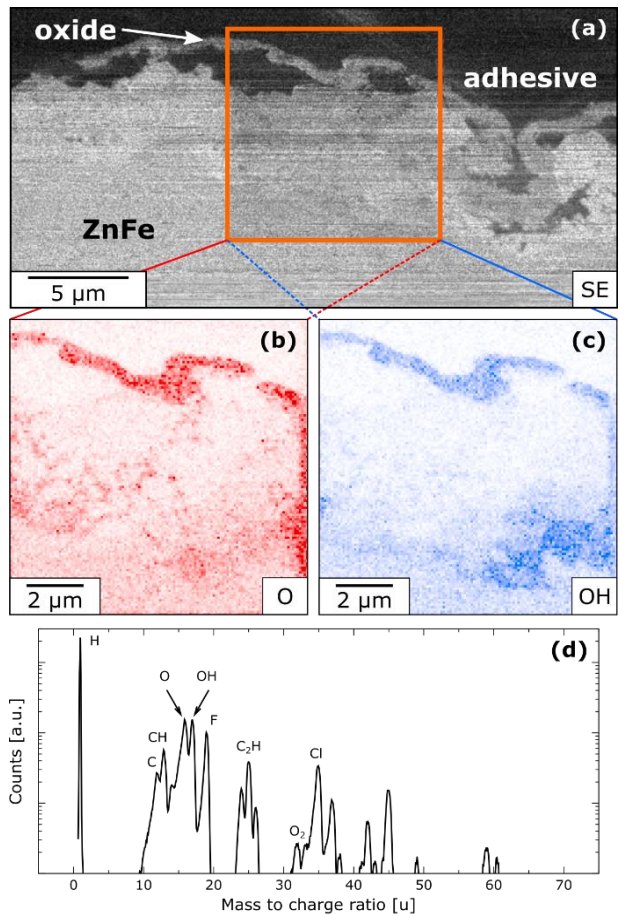

**Figure 6** Negatively biased ToF–SIMS measurements on a cross section polished and epoxy enhanced 22MnCrB8–2 + GA (200 s > 870 °C) sample with SE image (a), elemental mappings of O (b) and OH (c) and the sum spectrum of the measurement (d). The horizontal streaks in (a) are a result from a fast recording procedure.

### 3.4. Nanoscale TEM analysis

As the results from SEM and HIM have shown, further investigations on a higher magnification are necessary to allow a better description of the oxide layer, especially of the finely structured interface between Zn–Fe and oxide. During FIB preparation of the lamella, the focus of the thinning process was to ensure a homogeneously thin electron transparent window at the interface region between oxide and intermetallic phases. The shape of the final TEM–lamellas is influenced by residual strain within the material (see Fig. 7). As the sample becomes thinner, the tension gets released and results in small bulging and a slight distortion of the thinned lamella. The bulged material is cut away and therefore, the lamella gets thinner in this area, which is typically at the center.

As a result, the upper oxide layer is partially sputtered away and an increased curtaining effect can be observed as displayed in the micrograph in Fig. 7a. Different phases can be distinguished by the Z–contrast in the HAADF image. Within the oxide layer, one can identify two different materials. Between oxide and Zn–Fe phases, a thin separation layer can be observed. This layer is undamaged in the left half of the lamella, but appears to be fractured on the right half.

A more detailed view on the elemental composition is given by STEM–EDX analysis of the GA coated 45 s hardened 22MnCrB8–2 depicted in Fig. 7. The EDX mapping in Fig. 7b gives an overview of the elemental distributions in the sample. The oxide layer is divided in a large Mn–rich part and a second Zn–rich part. The thin film of Al and Cr oxide acts as a separation layer to the subjacent Zn–Fe phases. The interface consists of an $Al_2O_3$ layer with less than 100 nm in thickness, where some parts are fractured and replaced by $Cr_2O_3$.

A detailed EDX mapping with high magnification of the interface region in Fig. 7c gives an in–depth view of the oxide to Zn–Fe interface. The corresponding quantification of the sum spectra from the highlighted regions can be found in Tab. 2. The bottom Γ–phase (A) is clearly separated from the upper oxides. Zn–rich (B) and Mn–rich (C) oxides sit on top of $Al_2O_3$ (D), which is accompanied by $Cr_2O_3$. Moreover, the original $Al_2O_3$ layer is partially damaged in the left half of the recording and seems to be substituted by $Cr_2O_3$ (E).

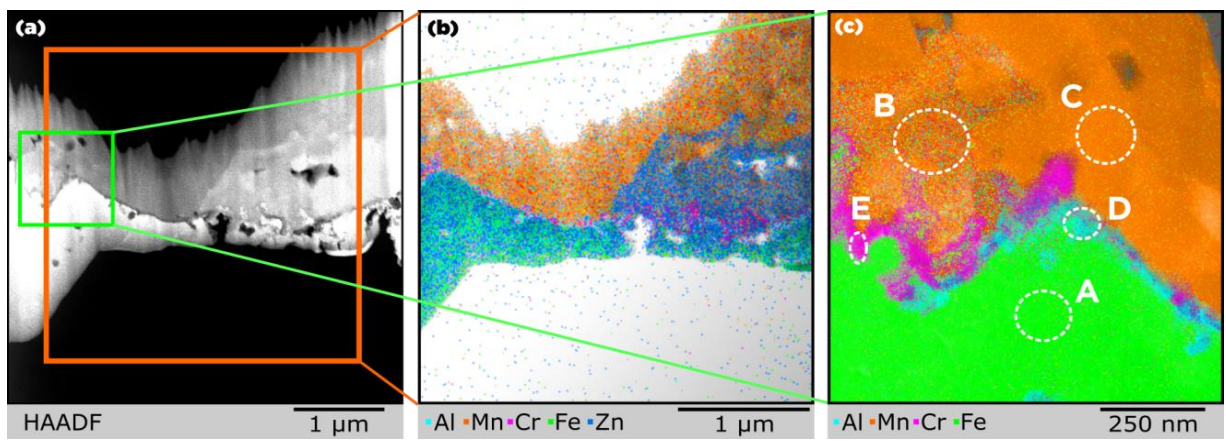

**Figure 7 TEM–lamella of 22MnCrB8–2 + GA (45 s > 870 °C) sample; (a) STEM–HAADF image reveals at least two phases in the oxide regime (top) separated by a thin layer from Zn–Fe intermetallic phases (bottom); (b) An overview EDX mapping separates the oxide phases into a Mn–rich and a Zn–rich oxide. (c) High magnification EDX mapping shows that the interface layer consists of $Al_2O_3$ and $Cr_2O_3$.**

**Table 2 Quantified EDX spectra of the highlighted areas in Fig. 7c. For the quantification, a mean thickness of 200 nm and density of 5,5 g/cm³ is used.**

| Area | Element concentration in at.% | | | | | |
|---|---|---|---|---|---|---|
| | O | Al | Cr | Mn | Fe | Zn |
| A | 27,2 | 0,1 | 0 | 0,3 | 14,2 | 58,0 |
| B | 72,4 | 0,1 | 0 | 4,0 | 0,2 | 23,1 |
| C | 73,6 | 0,0 | 0 | 20,8 | 0,2 | 5,2 |
| D | 66,7 | 15,2 | 2,1 | 3,5 | 1,9 | 10,4 |
| E | 68,2 | 0 | 12,7 | 2,0 | 1,8 | 15,1 |

Measurements for the 200 s hardened HX340LAD specimen with GI coating are depicted in Fig. 8. The overall structure is similar to the 45 s hardened samples, with a few remarkable differences. The STEM–HAADF image in Fig. 8a shows the elemental distribution very nicely with different gray values. For example, one can see the Mn–rich area within the oxide as a darker shade, compared to the dominant but brighter Zn area. The HAADF image also shows a thickness gradient in the thinned area, which indicates the lamella becoming thinner from top to bottom and from left and right into the center. The large overview EDX mapping in Fig. 8b confirms that the oxide layer consists of Mn and Zn oxides with grains of different sizes and orientation.

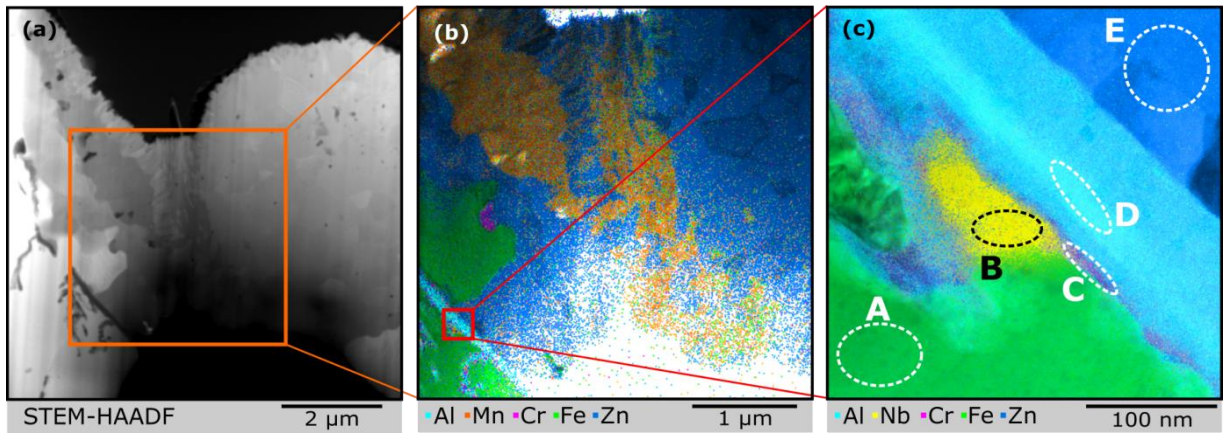

**Figure 8 TEM lamella of a HX340LAD + GI (200 s > 870 °C) sample; (a) STEM–HAADF image shows the multi–phase, polycrystalline oxide–layer (center) and Zn–Fe phase (bottom left); (b) Overview EDX mapping of the indicated area reveals two different oxide phases (Mn– or Zn– rich); (c) High magnification EDX mapping with focus on the initial $Al_2O_3$ layer.**

The EDX mapping also reveals a distinct layer of $Al_2O_3$, which lies clearly beneath the large oxide layer. On a closer look, one can see that the layer is fractured and integrated partially into the Zn–Fe phase. The Zn–Fe phase could escape the $Al_2O_3$ layer at a fractured area on the left–hand side. Along with Al, an increased concentration of Cr is present at the interface. If one looks at the $Al_2O_3$ layer at very high magnification (see Fig. 8c), a faint Cr enrichment at the bottom side of the $Al_2O_3$ is visible. Moreover, there is a small but clearly visible Nb–rich precipitate attached to the $(Al, Cr)_2O_3$ oxide layer.

For a better insight into the different elemental distributions of the involved phases, Table 3 shows different quantified areas as indicated in Fig. 8c. These values show an overall unexpectedly high amount of oxygen, which is due to difficulties of EDX with light elements. One can see partially oxidized Zn–Fe in (A). Nb oxide appears in (B) and is probably $NbO_2$ or $Nb_2O_5$ [26]. Chromium and aluminum appear both in the stable corundum structure as $(Al, Cr)_2O_3$ in (C) and (D) and Zn in (E) is native ZnO.

**Table 3 Quantified EDX spectra in the highlighted areas in Fig. 8c. For the quantification, a mean thickness of 200 nm and density of 5,5 g/cm³ is used.**

| Area | Element concentration in at.% | | | | | | |
|---|---|---|---|---|---|---|---|
| | O | Al | Cr | Mn | Fe | Zn | Nb |
| A | 29,2 | 0,5 | 0 | 0 | 50,3 | 19,8 | 0 |
| B | 84,0 | 0,9 | 0,2 | 4,5 | 1,0 | 1,1 | 8,1 |
| C | 71,3 | 11,6 | 5,7 | 0,6 | 1,8 | 7,4 | 1,5 |
| D | 70,0 | 28,0 | 0 | 0 | 0,2 | 1,7 | 0 |
| E | 72,8 | 0,7 | 0 | 1,6 | 0,6 | 24,9 | 0 |

### 3.5. Crystallographic analysis

Fig. 9a shows a TEM–BF image of an investigated Mn–rich oxide grain with the used selected area diffraction aperture fitting the grain. The resulting SAD pattern is depicted in Fig. 9b and the best match of diffraction pattern simulations has been found for tetragonal $ZnMn_2O_4$ spinel in [5,0,2] direction, as the simulated pattern in Fig. 9c confirms. The quantified STEM–EDX spectra measured in the highlighted areas A and B match the found spinel structure. However, Autengruber [3] found $Mn_3O_4$ on top of PHS instead of the here presented $ZnMn_2O_4$. Both minerals are very similar as they have the same crystal structure (space group $I4_1/amd$) and similar lattice constants (see Tab. 4), making a clear identification difficult. Furthermore, depending on the ambient conditions during oxide–formation, Zn and Mn may interchange within the spinel crystal structure and form $(Zn, Mn)Mn_2O_4$. According to EDX–data, Mn based spinel carries trace amounts of Fe, which can replace Zn or Mn as well.

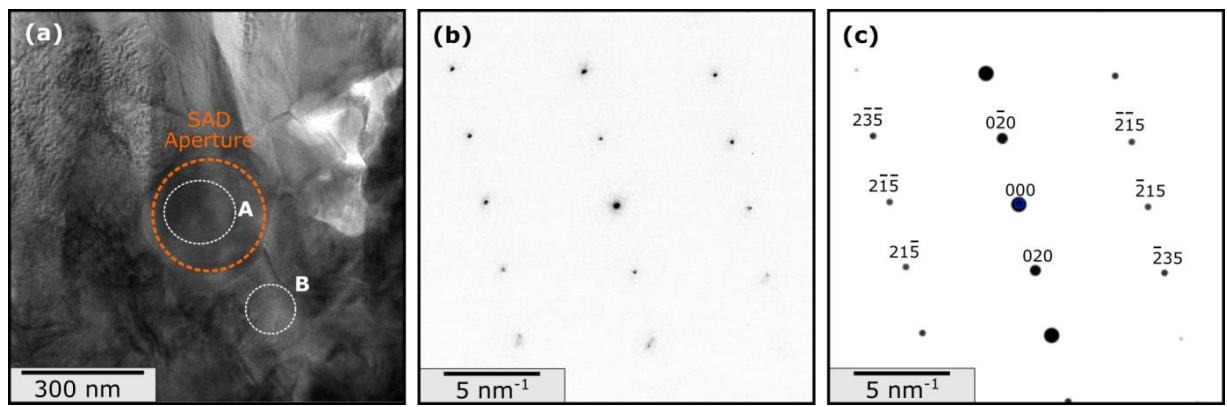

**Figure 9 TEM–SAD analysis of a single Mn–rich oxide grain on a 20MnB8 + GA (200 s > 870 °C) sample; The TEM–BF image (a) shows the applied SAD aperture and indicated EDX measurements as given by Tab. 4; The acquired diffraction pattern (b) fits the simulated diffraction pattern (c) for $ZnMn_2O_4$.**

**Table 4 Quantified EDX spectra in the highlighted areas in Fig. 9a.**

| Area / Mineral | Formula | Element concentration in at.% | | | | Crystal parameters | |
|---|---|---|---|---|---|---|---|
| | | O | Mn | Fe | Zn | a in Å | c in Å |
| A | - | 63,95 | 25,19 | 0,79 | 10,07 | - | - |
| B | - | 57,83 | 1,18 | 0,30 | 40,69 | - | - |
| Hetaerolite | $ZnMn_2O_4$ | 57,14 | 28,57 | - | 14,29 | 5,74 | 9,15 |
| Hausmannite | $Mn_3O_4$ | 57,14 | 42,86 | - | - | 5,76 | 9,44 |
| Zincite | ZnO | 50,00 | - | - | 50,00 | 3,25 | 5,21 |

## 4. Discussion

Based on our microscopy and spectroscopy characterization, we gained a number of insights on the role of steel alloying elements in the formation of specific oxides. Chemical analysis with SEM and AES on a µm scale left some oxide phases unidentified due to the limits of the spectroscopic techniques in the uppermost coating layer. Complementary, high magnification TEM investigations on nm–scale showed complex oxide formations along the interface between the oxide and the Zn–Fe intermetallic phases.

The main question is how the different alloy compositions influence the formation of the surface oxide layer during the annealing process. With a focus on the differences in the alloy and coating compositions, the investigated specimens can be divided into $2 \times 2$ groups. A first distinction can be based on the coating type (GI or GA). A second distinction can be made by respecting the alloy composition. HX340LAD and 20MnB8 are steel grades with no Cr as alloying element, while 22MnB5 and 22MnCrB8–2 contain low amounts of Cr.

The brittle ZnO layer is lifted off the intermetallic Zn–Fe phase creating large cavities with diameters of several µm. Bellhouse and McDermid [27] found that different thermal expansion coefficients of oxides and Fe are responsible for chipped off oxides during annealing and quenching on TRIP steel. Chen et al. [11] confirmed this behavior in a dual phase steel similar to the investigated steel grades. Due to the storage of the specimen in standard atmosphere, the Zn patina has already started to alter and form $Zn(OH)_2$ [28] as HIM–ToF–SIMS measurements confirm.

Aluminum is necessary as part of the galvanizing bath but not necessarily a desired element in the alloy composition. After hot–dipping, a faint $Al_2O_3$ layer is immediately formed on top of the Zn coating [13]. This protective film is present in all investigated specimens and separates the oxide layer from the intermetallic Zn–Fe phases. The $Al_2O_3$ is often heavily fractured and remnant clusters of small particles are incorporated into the intermetallic Γ–phase (see Fig. 10b). $Al_2O_3$ acts as a barrier, where precipitates of other alloying elements are captured. In direct comparison of GI and GA coatings, the noticeably thicker $Al_2O_3$ in GI coated specimens is a result of a higher Al content in the galvanizing bath. In areas where the whole oxide is lifted off the Zn–Fe phases, $Al_2O_3$ is mostly found at the bottom side of the oxide and not on top of the intermetallic Zn–Fe phases (see Fig. 10a,c).

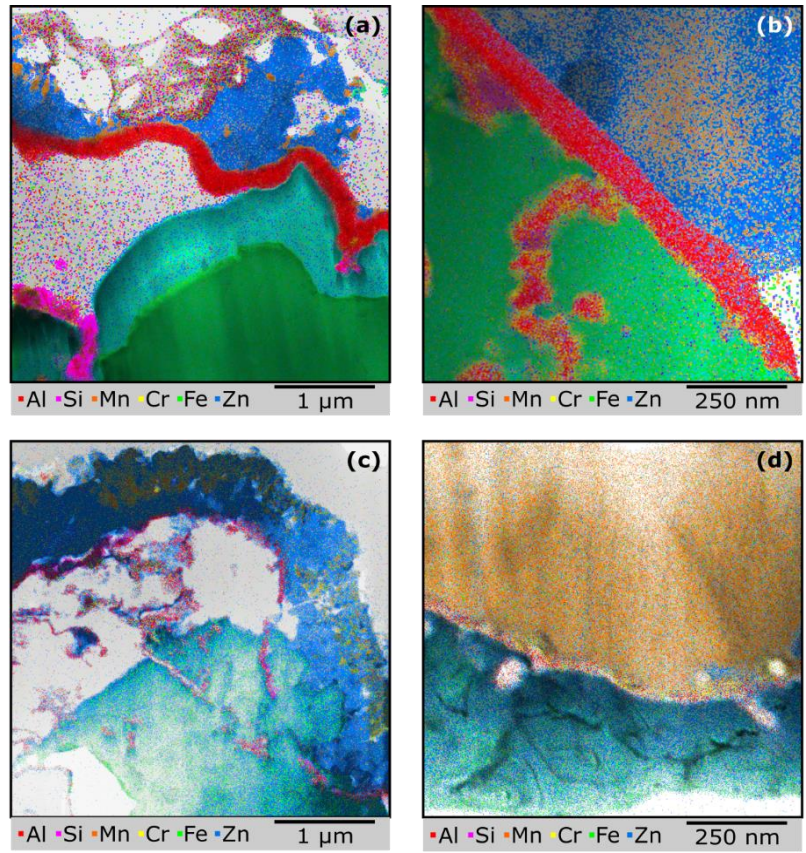

**Figure 10 Overlays of EDX maps on STEM–BF images for GI coated specimens 22MnB5 (a), HX340LAD (b) and GA coated specimens 22MnCrB8–2 (c), 20MnB8 (d).**

The main alloying elements of the investigated specimen is manganese, and thus the most common oxide–forming element beside Zn and Al. Mn is used to alter the austenitization condition of the steel. EDX measurements combined with SAD revealed that Mn does not occur in its simple native oxide forms MnO, $Mn_2O_3$ or $MnO_2$ but in form of a mixed spinel $(Mn,Zn)Mn_2O_4$. Autengruber et al. [13] already found the spinel to be hausmannite $Mn_3O_4$ (MnO + $Mn_2O_3$), but our measurements suggest that MnO can be substituted by ZnO within the spinel crystal at any time, forming the slightly different $ZnMn_2O_4$ spinel, a mineral called hetaerolite. The measurements show a small but noticeable amount of Fe in the spinel grains, hinting at an additional replacement of Mn by Fe [29]. However, the concentrations are low and occasionally not detectable at all. The spinel oxide is found in the main oxide layer above the $Al_2O_3$ separation and is predominantly embedded in ZnO (see Fig. 10a–d).

A common addition in hot–forming steel alloys is chromium, which further changes the austenitization condition similarly to Mn. In the bare steel sheet, Cr forms an oxide acting as corrosion protection close to the surface. Its most common native oxide is $Cr_2O_3$ in corundum form, which is also the only oxide stable at the annealing temperature of 890 °C [30,31]. Our previous measurements suggest a high affinity of $Cr_2O_3$ to the original $Al_2O_3$ layer. During annealing, $Cr_2O_3$ particles seem to attach onto the steel side of the $Al_2O_3$ layer, forming an additional thin diffusion and oxidation barrier (see Fig. 10b,d). This behavior can be observed for high Cr alloyed steels but to a lesser degree also in low Cr alloyed steels. In sufficient concentrations, the $Cr_2O_3$ layer can act as a replacement at the interface between oxide and the Zn–Fe intermetallic phases. At this interface the damaged $Al_2O_3$ layer is supplemented with $Cr_2O_3$, repairing the barrier.

Although silicon occurs only in low concentrations as an alloying element, $SiO_2$ is a common product on top of the Zn coated steel sheets. Similar to Cr, Si acts as a deoxidizer in the steel matrix. Due to the very high stability of $SiO_2$, the heat treatment has no effect on the oxide and precipitates can move freely in liquefied Zn(Fe) coatings. Eventually, these precipitates will appear on top of the intermetallic Zn–Fe phases and form small structures underneath the $Al_2O_3$ separation layer (see Fig. 10a).

Other low content additions like niobium can be found as sub µm sized oxide particles trapped at the $Al_2O_3$ layer. It can be assumed that the Nb oxide is the most common compound $Nb_2O_5$, which is thermodynamically stable below 1512 °C [32]. Due to the very low concentration of Nb in the steel alloys, precipitates are only detected in rare cases.

HIM–ToF–SIMS investigations on GA coated specimen revealed the coexistence of oxides and hydroxides as part of the uppermost layer as predicted by Lindström and Wallinder [33]. Moreover, a locally restricted appearance of OH phases (see Fig. 6c) can be seen underneath the uppermost coating layer, suggesting that conversion of ZnO to $Zn(OH)_2$ can take place underneath the surface. Because the evolution from oxide to hydroxide can only happen if $H_2O$ is available, we conclude that water droplets can penetrate the upper oxide layer through cracks and accelerate the conversion process significantly.

Based on the presented observations and assumptions, we developed a schematic model to describe the oxide formation (see Fig. 11). The model is explained with focus on GI coated specimens, but can be easily adopted for GA coated samples, where the initial structure is similar to a fully δ–phase transitioned Stage II instead of Stage I.

Before heating up in the press hardening furnace (Stage I), the layered structure consists of the steel substrate, containing alloying elements like Si, Cr or Mn and the coating consisting of Al and Zn. Due to the proximity to the steel–sheet surface, these elements are mainly oxides as a result from selective oxidation from the hot–dip galvanizing annealing process. The steel matrix is separated from the Zn coating by a thin $Fe_2Al_5$ inhibition layer, which acts as a diffusion barrier. Due to the low Al additions in the liquid Zn bath and the high affinity of Al to O, a several nanometers thick $Al_2O_3$ layer covers the coating.

Stage II depicts the diffusion of Fe and alloying elements into the Zn coating, either through galvannealing (in the hot–dip galvanizing process) or during heating up in the press hardening furnace to about 550 °C. This results in a transformation of Fe and Zn into intermetallic Zn–Fe phases. In most cases liquid Zn is completely transformed into solid Zn–Fe phases also deforming the outermost surface. During the phase transformation, small alloy particles can move through the whole coating, eventually reaching the $Al_2O_3$ layer.

After the specimen reaches a temperature of 665 °C, a transition from δ– to Γ–phase happens (Stage III). A consequence of the phase transformation and the heating rate is a volume change of the coating, deforming the $Al_2O_3$ layer [14]. If the induced strain on the oxide is too high, it will break apart and can be incorporated in the Zn–Fe phase.

At the final annealing temperature of 890 °C (Stage IV), the Zn–Fe phase is already decomposed at 780 °C into Zn saturated α–Fe and liquid Zn. The α–Fe grains start to grow from the steel substrate toward the surface and push the alloying elements and their oxides in the same direction. The liquid Zn can protrude through the broken $Al_2O_3$ layer and will react with ambient O to form ZnO. The Mn containing liquid Zn is transported above the $Al_2O_3$ layer. There, Mn and Zn react with oxygen to form the spinel $(Mn,Zn)Mn_2O_4$, depending on the availability of Mn. When Mn is depleted in the fluid, another layer of ZnO is typically formed, covering most of the spinel. The high thermodynamic stability and low density of $SiO_2$ lead to an expulsion from the liquefied Zn coating. Therefore, $SiO_2$ can be found on top of the final Zn–Fe intermetallic phases but will not surpass the $Al_2O_3$ layer. $Cr_2O_3$ is extremely stable with a melting point of 2435 °C, and thus is unaffected by the annealing process. $Cr_2O_3$ shares the same corundum crystal structure with $Al_2O_3$, explaining the high attraction of $Cr_2O_3$ to the initial $Al_2O_3$ layer. If this layer is unharmed, $Cr_2O_3$ is accumulated on the steel side, increasing and strengthening the former pure $Al_2O_3$ layer. But if fracturing happened during the previous stages and Cr is available in sufficient concentration, $Cr_2O_3$ can regenerate the interface by replacing missing $Al_2O_3$. A similar behavior was found in Fe–20Cr–25Ni–Nb austenitic stainless steel by Chen et al. [34], although the initial oxide was $Cr_2O_3$ instead of $Al_2O_3$.

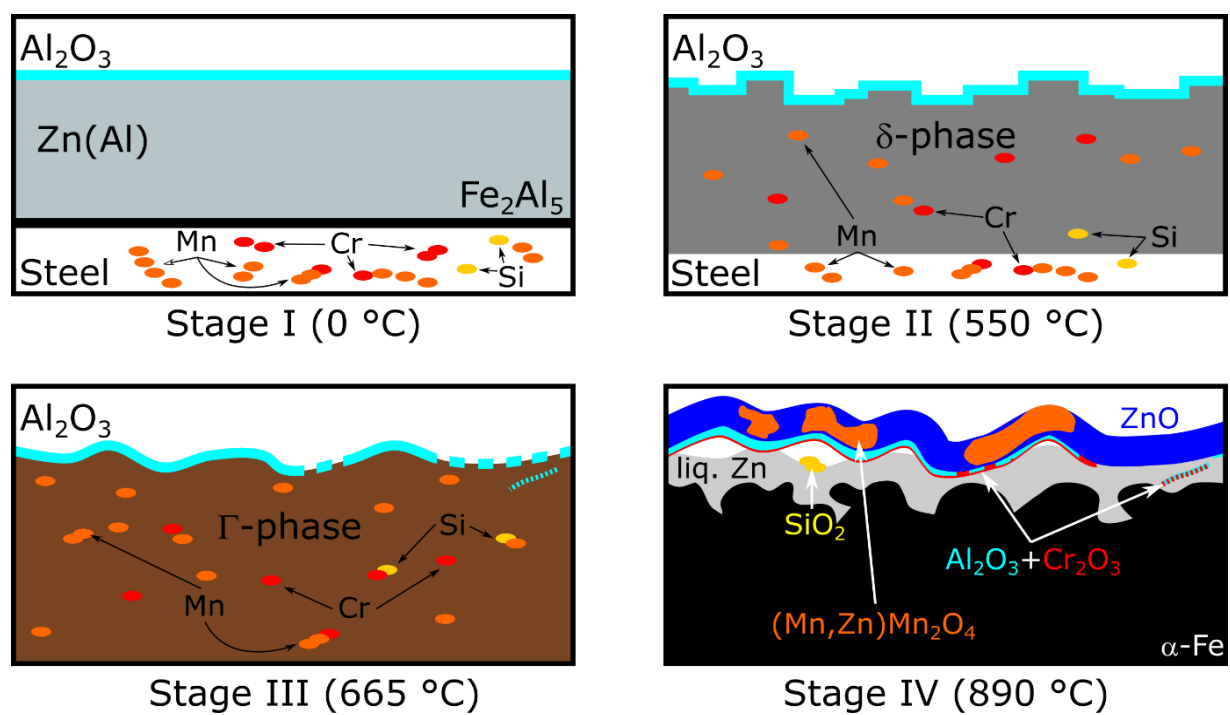

**Figure 11 Schematic of four important oxide–formation stages during austenitization annealing showing only dominant Zn–Fe phases.**

## 5. Conclusion

In this work, the oxide composition of eight different Zn–Fe coated steel sheets with either 45 s or 200 s austenitization holding time were analyzed by AES and EDX in a SEM, ToF–SIMS in a HIM and EDX and SAD in a (S)TEM. All specimens were hot–dip galvanized and half of them were additionally galvannealed. The used techniques allowed us to investigate the oxide distribution on a range from several µm down to a few nm.

The presented results show that the main part of the oxide layer consists of ZnO, which eventually will transform to $Zn(OH)_2$ over time in ambient atmosphere. The dominant polycrystalline ZnO layer is accompanied by a spinel oxide $(Mn,Zn)Mn_2O_4$, where Mn and Zn can substitute each other. Additionally, low amounts of Fe from the steel substrate are found along the spinel phases. The primary oxide layer is usually separated by a thin $Al_2O_3$ film, originating from low Al additions in the Zn bath during hot–dip galvanization. The main oxide layer is often lifted off the subjacent intermetallic Zn–Fe phases, where the faint $Al_2O_3$ layer is typically attached to the bottom side of the oxide layer but not onto the intermetallic Zn–Fe phases. Additionally, phase transformations during austenitization annealing leads to volume changes, which can damage the otherwise closed $Al_2O_3$ coating. Remnants from these cracks are clusters of small $Al_2O_3$ particles and can be found in the Γ–phase. If Cr is available from the steel alloy, it will act as an addition and enhancement for Al, attaching to the $Al_2O_3$ layer and oxidizing to $Cr_2O_3$. As Cr typically is not found on the surface, it can be assumed that it was already oxidized before annealing (e.g. by selective oxidation). During annealing, the $(Al,Cr)_2O_3$ layer behaves like a filter in the liquid Zn, which allows only Zn and Mn to pass through and form oxides above, but traps precipitates of other elements like Nb.

## Acknowledgement

The financial support by the Austrian Federal Ministry for Digital and Economic Affairs and the National Foundation for Research, Technology and Development in the frame of the CDL for Nanoscale Phase Transformations is gratefully acknowledged.